\documentstyle[12pt,aasms4]{article}
\begin{document}
\title{THE UNIQUE RAPID VARIABILITIES OF THE IRON K$\alpha$ LINE PROFILES 
IN NGC 4151}
\author{Jun-Xian Wang$^{1,2}$, Ting-Gui Wang$^{1,2}$, and You-Yuan
Zhou$^{1,2,3}$}
\altaffiltext{1}{Center for Astrophysics,University of Science and
     Technology of China, Hefei, Anhui, 230026, P. R. China;
     jxw@mail.ustc.edu.cn}
\altaffiltext{2}{National Astronomical Observatories, Chinese Academy of
     Sciences}
\altaffiltext{3}{Beijing Astrophysics Center, Beijing, 100080, P. R.
     China}

\begin{abstract}

We present a detailed analysis of the iron K$\alpha$ line variabilities
in NGC 4151 by using long ASCA observation data obtained in May 1995.
Despite the relatively small amplitude variations in the continuum flux,
the iron K$\alpha$ line flux and profile show dramatic variations.
Particularly, the line profile changes from single peak to seeming
double peaks and back in time scales of a few 10$^4$ sec. 
The seemingly double-peaked profiles can be well interpreted as line 
emission from a Keplerian ring around a massive black hole.
An absorption line at around 5.9 keV is also marginnaly detected.
We discussed current Fe K line models, but none of them can well explain the 
observed line and continuum variations.
\end{abstract}
\keywords{black hole physics --- galaxies: active --- galaxies: individual
(NGC 4151) --- line: profiles --- X-rays: galaxies}

\section{Introduction}
Broad Fe K fluorescence line profiles, detected in many AGNs
(Tanaka et al. 1995, Nandra et al. 1997, Wang et al. 1999a, Nandra et al.
1999, and references therein), are thought to arise from the
accretion disk near the center massive black hole of AGNs.
The line profile, thus, carries important information about the distribution 
and kinematics of optically material as well as about the space properties in 
the vicinity of the postulated super massive black hole. 
Doppler and gravitational shifts would imprint
characteristic signatures on the line profile which map the geometric and
dynamical distributions of matter surround the black hole.

Additional information
concerning the geometry of X-ray source and matter in the active nucleus
could in principle be derived by studying the rapid variability (on 
time scales of several 10$^4$ s or less) of the
line profile, intensity and their relationship with the continuum
variations.
So far, rapid variability in Iron K line has been detected in NGC 7314
(Yaqoob et al. 1996), MCG --6-30-15 (Iwasawa et al. 1996, 1999), NGC 4051
(Wang et al. 1999b) and NGC 3516 (Nandra et al. 1999).

The observed Fe K$\alpha$ line rapid variabilities in different targets
or during different observations of the same target are quite different.
Yaqoob el al. (1996) presented the evidence for rapid variability of
the Fe K line profile in the narrow-line Seyfert galaxy NGC 7314, which is
consistent with a disk-line of constant equivalent width superposed on a
constant flux narrow line (presumably from the torus).
Iwasawa et al. (1996) discovered that, during the ASCA observation on
MCG --6-30-15 in 1994, when the source is bright, the Fe K line is
weak and dominated by the narrow core, whilst during a deep minimum,
a huge red tail appears. The intensity of broad Fe K line correlates
inversely with the continuum flux on the time scales of several 10$^4$ s.
And Iwasawa et al. (1999) discovered that during the ASCA observation
on the same target in 1997, the Fe K$\alpha$ line of a flare phase
peaks around 5keV and most of its emission is shifted to below 6 keV with no
component detected at 6.4keV.
While for NGC 4051, during the ASCA observation in 1994, the equivalent
width and the width of Fe K line correlate positively with the continuum
flux (Wang et al. 1999b), which shows an opposite trend with the ASCA
observation on MCG --6-30-15 in 1994.
The rapid variabilities of Fe K$\alpha$ line profiles in NGC 3516 are
also quite irregular (Nandra et al. 1999, Wang et al. 2000).

The nearby Seyfert 1.5 galaxy (z = 0.0033), NGC 4151, was first 
established as a bright X-ray source some thirty years ago (Gursky et al.
1971). Since that time, this archetypal low-luminosity active galaxy has
invariably been considered as a prime target by all
major X-ray astronomy missions. 
In 1995, NGC 4151 was observed by ASCA from May 10 to 12 (Leighly et al.
1997). The net exposure time of the observation is about 200 ks.
Because of the enhanced sensitivity provided by ASCA and the much longer
exposure time, the obtained Fe K$\alpha$ line profile has very high data
quality (Wang et al. 1999a, here after W99). 
In this letter we report the discovery of unique and rapid
variabilities of the iron line profiles in NGC 4151 during the ASCA observation
in 1995.
 
\section{Fitting the X-ray continuum} 
Only the SIS data are used in this paper to study the Fe K$\alpha$ line  
variabilities. Detailed description of the ASCA observation and data reduction 
can be found in W99.
Figure 1a shows the background subtracted light curve (0.4 -- 10.0 keV)
with 256 s binning. Significant variability
is absent on time scales of less than 1000s, but moderate variations can be
seen on time scales of several $10^4$ s, exhibiting a continuous slow
drifting behavior (e.g. Lawrence 1980).
The light curve shows an intensity dip at 60 ks from the beginning, which
lasts about 50 ks, and two neighboring flares
at about 150 ks, for which each lasts about 30 ks. The intensity ratios of
the flares to the dip are around 0.8.
Apart from these, the count rate
also increases slightly during the whole observation. 

The time-averaged Fe K$\alpha$ line profile from this observation has been
extracted by W99. The data quality of the line profile is one of the best
that could be found in literatures (see Figure 1a in W99).
In order to analysis the rapid variability properties of the Fe K$\alpha$
line profiles,  we divide the whole observation into five time intervals
(see Figure 1, I-1 to I-5). Following Weaver et al. (1994) and W99, we fit
the underlying continuum of the each segment in
the 1.0 -- 4.0 and 8.0 -- 10.0 keV bands (to exclude the broad iron  
line region) with a model which consists of a dual absorbed power law plus
some fraction (the best-fit value for the time averaged spectrum is $\sim$ 5\%) 
of the direct continuum scattered into our line of sight and
absorbed only by the Galactic column of 2 $\times$ 10$^{20}$ cm$^{-2}$.
We do not include a Compton reflection component in the spectral fits
because of its small impact on the ASCA spectra. 
During the fitting, we fix the power law index and the flux of the scattered
component at their best-fit values obtained
from the time-averaged spectrum. Same to the time-averaged spectrum, the
spectra of these five segments can also be well fitted by this model.
We show the results in Table 1. 

The power law indices for individual segments are all well consistent 
with the average value, i.e, there is no significant change in the continuum
spectrum. However, the absorption columns density and the covering factor
of the dual absorber showed significant variations (table 1), not
correlated with variations of the continuum flux.

\section{The Fe K$\alpha$ line profiles} 
Figure 2 shows the Fe K$\alpha$ line profiles for the whole observation and
the five time intervals. The profiles are obtained from the ratio, 
the data divided by the best-fit continuum (folded through the detector
response), multiplied by the power-law (original form), thus are detector
effective area corrected, and also independent from the model used to fitting
the lines. The time-averaged line shows a strong narrow
peak around 6.4 keV and a huge red wing extending to $\sim$ 4.5 keV.
Apart from a similar narrow core in the 6.4 keV, the line profile 
apparently differs by one interval to the other.  
The line profile of I-1 also shows a huge red wing, extending
to about 4.0 keV with a possible red peak at around 4.9 keV; while for
I-2, the red wing is much weaker.
The line profile of I-5 is quite similar to the time averaged one.
For I-3, besides the strong peak at 6.4 keV, there is also a seeming
peak at around 5.4 keV. 
We noticed that between the two peaks there seems a valley in between
which indicates that the line is double-peaked.
The line profile of I-4 is similar to that of I-3, seemingly double-peaked, 
but much weaker.
We conclude here that the Fe K$\alpha$ line profiles of NGC 4151 show
clear and obvious variations during the ASCA observation.

Initially the line profiles are fitted with two gaussians and the
results is given in Table 2. A narrow gaussian at 
around 6.4 keV and additional broad component can well fit the line profiles.
and the results clearly present the variations of the line profiles and 
equivalent widths.

We also fit the Fe K line profiles with a model consist of a narrow component 
and a disk line (Fabian et al. 1989). The narrow core is likely emitted  
by material far away from the center black hole, such as torus, thus we 
would not expect substantial change in the line flux within the duration of the 
observation.  During the fitting, we fix the flux
of the narrow core for the five time intervals at the best-fit value from
the time-averaged spectrum (which has a equivalent width of 70 eV according
to the time-averaged spectrum). We also fix the disk-line energy at 6.4 keV in 
the source rest frame (W99), and fix the inclination angles of the disk at
the best-fit value from the time-averaged spectrum. In fact, when
we free the inclinations for the five time intervals, the best-fit values are
all well consistent with that of the time-averaged spectrum (27 degree).
As W99 did, the corresponding Fe K$\beta$ and Ni K$\alpha$ components 
(disk-line plus narrow core, see W99 for detail) are also included
in the fit. The model can fit the line profiles quite well as
the two Gaussians model do, and the results of the fit are shown in Table 3.
We did not adopt the Model B in W99 because the scattered disk line
component should be constant on such a short time scale, which is 
inconsistent with the rapid Fe K line variabilities we detected.

Nandra et al. (1999) claimed that they discovered a red shifted iron
resonant absorption line at around 5.9 keV in the Fe K$\alpha$ profile of
NGC 3516. We also add an additional absorption line component to the
model, and find that including such a line can significantly improve the
fit for I-4 ($\delta\chi^2$ = 7). However, the absorption line is not
required for other segments ($\delta\chi^2$ $<$ 3.0).

\section{Discussion}

\subsection{The seemingly double-peaked line profiles}

Massive black holes are generally thought to exist at the centers of
active galaxies (Rees 1984), but unambiguous identification of a black
hole has been impeded by lack of evidence for the strong-field
relativistic effects expected in the vicinity of black holes.
Because of its unique anticipated profiles, the Fe K$\alpha$
fluorescence line has long been considered as an observational key to the
study of the innermost regions of active galactic nuclei (AGN).
Fabian et al. (1989) and Laor (1991) pointed out that the line
profiles from the accretion disk should be broad, skewed and
double-peaked (dependent on the inclination of the disk and the
line-producing region in the disk), which are characteristics
of the Doppler effects of the accretion disk and the strong
gravitational field in the vicinity of the central black hole.

The Fe K$\alpha$ line profiles detected in many AGNs have been proved to be
skewed and extremely broad (Tanaka et al. 1995, Nandra et al. 1997, Wang
et al. 1999a, Nandra et al. 1999, and references therein), consistent with
the disk line model, so give the ever strongest evidence for the presence
of massive black holes and accretion disks in the center of AGNs.

However, some alternative models, such as Comptonization and jet,
can also explain the broad and skewed Fe K$\alpha$ line profiles.
Those models do not require strong gravitational field and so give 
no support to the presence of
massive black holes (Fabian et al. 1995, Misra \& Sutaria 1999).
But different to the disk line model, those models predict a smooth line
feature, but not a double-peaked one. Though some indirect evidences have
shown that they are not satisfactory
(Fabian et al. 1995, Misra 2000, Reynolds \& Wilms 2000),
as Misra \& Kembhavi (1998) have pointed out, we infer that the
detection of a double-peaked line profile is the key to
confirm the disk line model for the observed Fe K$\alpha$ lines,
and so on confirm the detection of extremely strong gravitational field.

In this paper we find that the Fe K$\alpha$ line profile of NGC 4151
is seemingly double-peaked during phase I-3 and I-4 in Fig1. Because
an additional resonant absorption line at 5.9 keV has also been detected
in the line profile in I-4, we only focus on the profile
of I-3 in this section. We show the
seemingly double-peaked line profile of I-3 in detail in Fig.3, and the
best-fit two gaussians and disk-line models are also plotted.

The seemingly double-peaked line profile strongly suggest that the line
arises from materials with disk-like geometry. Additional strong 
gravitational field is also needed to explain the red shift of the profile. 
Others models for the broad line profiles, such as Comptonization
or jet, can not be responsible for a double-peaked line profile, and
disk-like outflows or inflows with high velocity can not be responsible 
for the red shift of the line profile. 
The best disk line fits show that the line mainly
arises from a narrow ring between 10.8 R$_g$ and 18.6 R$_g$.

Currently, perhaps due to the limited energy resolution and data quality,
obvious double-peaked Fe K$\alpha$ line profile has never been detected
in any active galactic nuclei. We noticed that the line profile of
MCG --6-30-15 in Fig. 1b of Wang et el. (1999a) looks double-peaked,
but the plot was folded through the effective area of the detectors which rises
sharply towards the low energy, so did not give the true profile of the line.
While in plottings corrected for detectors' effective area,
we can not see obvious double peaks (see Fig. 7 of Iwasawa
et al. 1996 and Fig. 6 of Fabian et al. 2000).
We hope that the new generation X-ray satellites, such as XMM, can
show us clearly double-peaked iron line profiles of AGNs.

\subsection{The rapid variabilities of the Fe K$\alpha$ line profiles}

Additional to the double-peaked line profiles, the rapid variabilities
of the Fe K line profiles in NGC 4151 are also detected in this letter.
We divided the whole ASCA observation of NGC 4151 in 1995 into five time
intervals, and found that, despite the relatively small amplitude 
variations in the continuum flux,
the Fe K$\alpha$ line flux and profiles show dramatic variations:
during the first time interval I-1, the iron
line profile is broad and skewed, showing a strong peak at 6.4 keV, and a
huge red wing extending to $\sim$ 4.5 keV, similar to the time-averaged
line profile; after several 10$^4$ s, the huge red wing becomes very weak
(I-2); for I-3, except the strong peak at 6.4 keV, there is also a seeming
peak at around 5.4 keV; the line profile of I-4 is similar to that of I-3,
but much weaker; and finally, the line profile change back to single
peak (I-5).  The line equivalent widths of the five intervals are 
375, 148, 329, 168 and 300 eV respectively, showing obvious variations.

The rapid variabilities of Fe K line profiles discovered previously
have suggested that the X-ray continuum flux is generated by magnetic
flares above the accretion disk, and at any given time there are only a
few flares moving around above the disk
(Iwasawa et al. 1996, Fabian 1997, Wang et al. 1999b, Iwasawa et al. 1999, 
Wang et al. 2000). Flares with different location and different motion can
give different fluorescence line profiles. 

Can the magnetic flare model also explain the observed
rapid variabilities of Fe K$\alpha$ line in NGC 4151?
For I-1 and I-5, the X-ray continuum (in the flare model context) could be 
dominated by a single 
or several strong flares located very close to the center black hole, while 
the inner disk radius from the disk line fits (7.4 and 8.3 R$_g$) gives 
strong supports to this model.
For I-2, the X-ray continuum seems to be dominated by flares
far away from the center black hole (R$_i$ = 17 R$_g$).  
For I-3 and I-4, the X-ray continuum could be dominated by a single or
several strong flares, which are  $\sim r_g$ above and coroating with
the disk at around 14 R$_g$, with only the materials in the ring around 
14 R$_g$ being illuminated.  The large dramatic decreasing in the line EWs 
from I-3 to I-4 might be due to different degree of the anisotropy in the 
X-ray continuum, i.e., larger fraction of the X-ray continuum received by the 
disk in I-3 than in I-4. However, it requires some fine tune in order to keep 
the observed continuum more or less constant. Alternatively,  
Misra (2000) suggested that the changes in
the disk structure may explain the observed line profile variations, but,
a disk undergoing structure changes is also hard to keep the same radiative 
spectrum. Finally, the decreasing of the fluorescent yielding may be due to the 
increasing ionization in the disk skin.  Under flare context, large change 
in the ionization structure of disk skin is possible even for quite moderate 
variations in the observed continuum flux if the flares have different heights 
If this is case, we would expect that the flare is lower in the I-4 
than in I-3. This should also reflect in the line profile. However, the 
uncertainty in the disk parameters do not allow us to verify this.  

The observed properties strongly suggest that the rapid line variations
may not be due the variations of X-ray continuum source.
Instead, it may be due to changes in geometry or ionization level of the
reflecting medium and at the same time, these changes should not affect
the X-ray producing medium. If so, except for the variations of the X-ray
continuum source, what can lead to the changes in geometry or ionization
level of the reflecting medium? 

An absorption line at 5.8 keV is marginally detected in I-4, but not 
significant in other intervals. Note a variable absorption line at the 
same energy has been discovered by Nandra et al. (1999) in NGC 3516. 
Nandra et al. (1999) interpreted this line as the redshifted Fe K 
resonant absorption line. The large redshift is either caused by the 
infalling absorbing material or by the strong gravity in the 
vicinity of the black hole. Given the fact that the line is highly 
variable and the large redshift, the agreement of the redshifts in 
this two objects perhaps is not purely coincidence, but have some 
physical origin. If so, the gravitational redshift explanation would 
be favored.  

\acknowledgments
This work is supported by Chinese National Natural Science Foundation,
Chinese Science and Technology ministry and Foundation of Ministry of
Education.

\newpage 
\input psfig.sty
\begin{figure}
\centering
\psfig{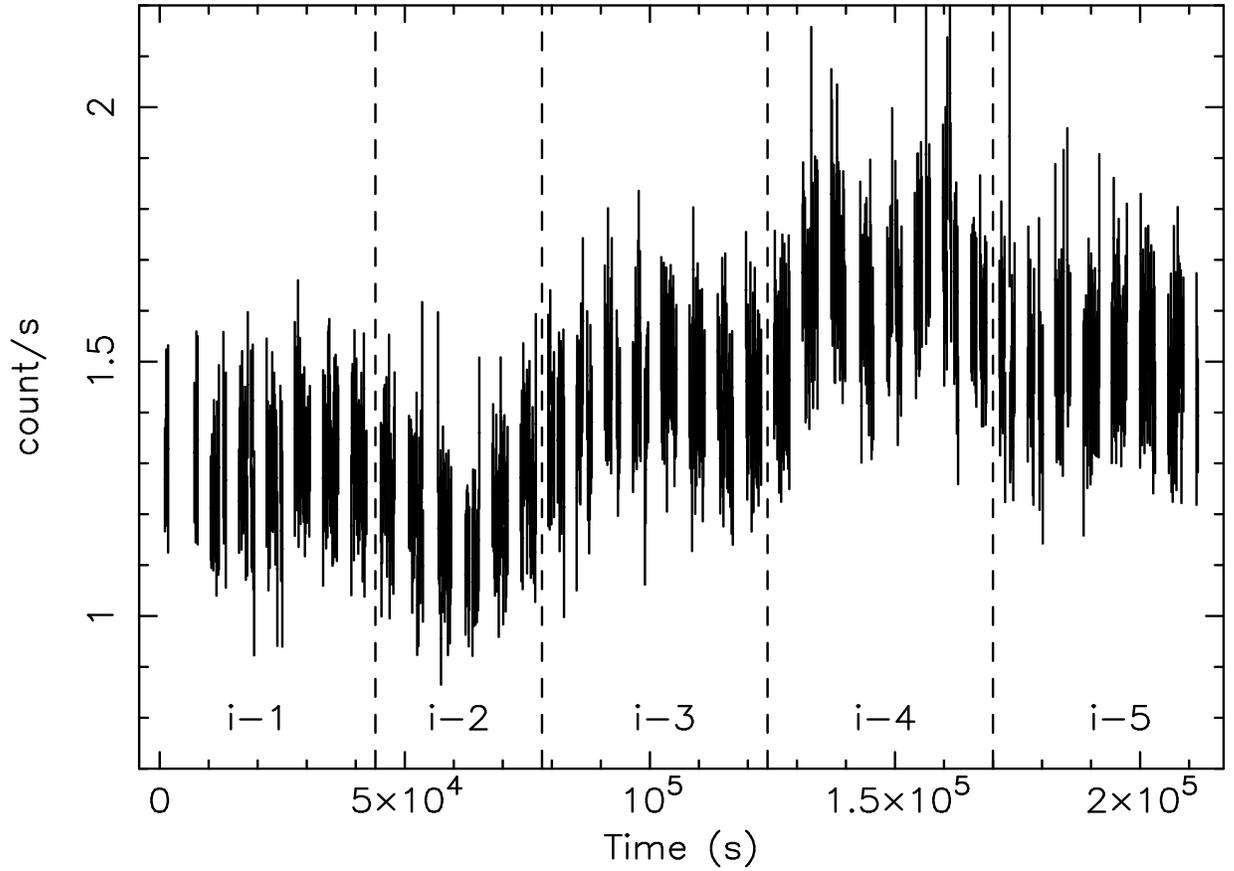}
\figcaption[f1.ps]{Background-subtracted light curve (0.4 -- 10.0 keV,
SIS) of NGC 4151 for the long ASCA observation taken in May 1995. The
light curve is binned using 256 s time intervals. The background were
estimated from source-free regions. The five time intervals are also
shown here. \label{fig-1}}
\end{figure}

\clearpage
\newpage
\begin{figure}
\psfig{figure=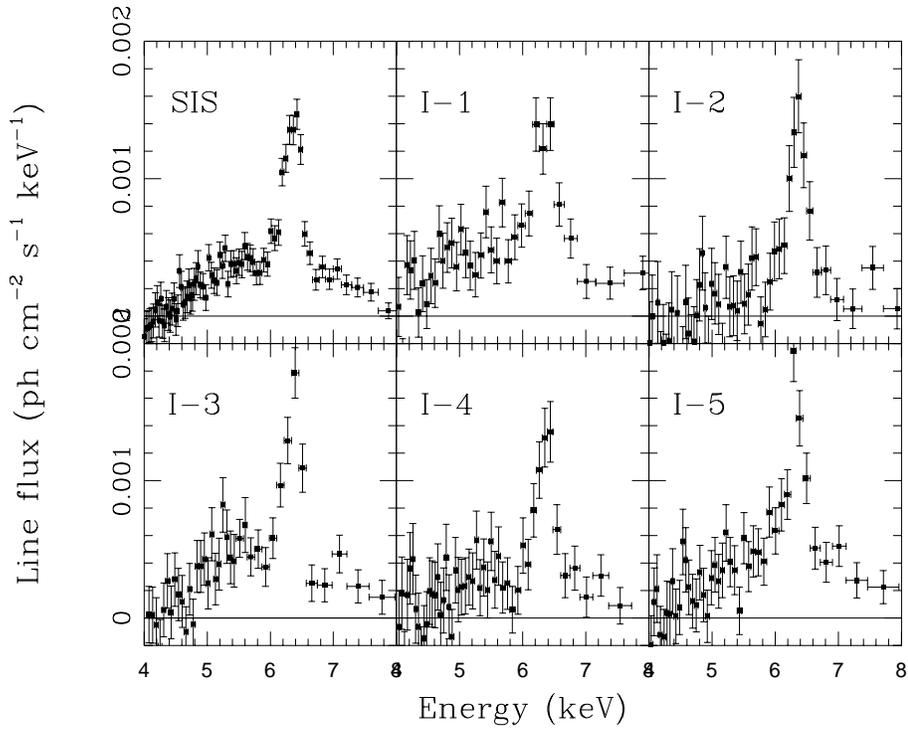,width=16.2cm,height=12.0cm,angle=-90}
\caption[f2.ps]{The Fe K$\alpha$ line profiles for the five time
intervals. see text for details.
\label{fig-2}}
\end{figure}

\clearpage
\newpage
\begin{figure}
\psfig{figure=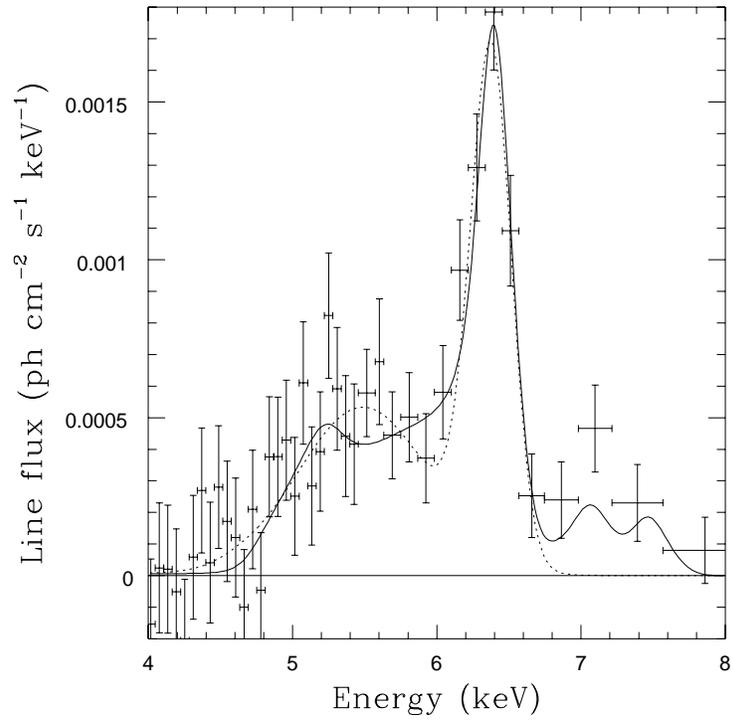,width=16.2cm,height=12.0cm,angle=-90}
\figcaption[f3.ps]{The line profile of I-3. dotted line: best-fit two 
gaussians; solid line: best-fit disk-line model.
\label{fig-3}}
\end{figure}

\newpage

\begin{table}
\caption[]{Dual absorbed power law fits for the 1.0 -- 10.0 keV continuum
of the five time intervals}
\begin{tabular}{ccccccccc} 
\\ \hline \hline
intervals(ks) &  NH1 & NH2 & cover factor & index & flux$^a$
& $\chi^2$/dof \\\hline

0. - end & 3.17$_{-0.15}^{+0.15}$ & 7.44$_{-0.99}^{+1.03}$ & 
0.62$_{-0.05}^{+0.04}$ & 1.34$_{-0.07}^{+0.06}$ & 2.292 & 638/570 \\

0. - 42  & 3.29$_{-0.60}^{+0.50}$ & 7.92$_{-2.97}^{+3.95}$ &
0.60$_{-0.11}^{+0.10}$  & 1.23$_{-0.19}^{+0.22}$ & 2.123 & 457/441 \\

42. - 75. & 3.26$_{-0.51}^{+0.49}$ & 10.17$_{-2.73}^{+3.32}$ &
0.73$_{-0.07}^{+0.05}$ & 1.50$_{-0.22}^{+0.25}$ &
2.322 & 485/434 \\

75. - 122. & 3.06$_{-0.51}^{+0.45}$ & 6.73$_{-1.61}^{+2.14}$ & 
0.69$_{-0.08}^{+0.08}$ & 1.36$_{-0.14}^{+0.15}$ & 2.424 & 477/452
\\

122. - 168. & 3.43$_{-0.37}^{+0.33}$ & 10.06$_{-4.39}^{+4.46}$ & 
0.49$_{-0.12}^{+0.11}$ & 1.34$_{-0.19}^{+0.24}$ & 2.810 & 415/456\\

168. - end & 2.43$_{-0.39}^{+0.37}$ & 6.06$_{-1.19}^{+1.48}$ & 
0.73$_{-0.06}^{+0.06}$ & 1.31$_{-0.13}^{+0.13}$ & 2.409 & 411/449\\
\hline
\\
\end{tabular}
$^a$10$^{-10}$erg/cm$^2$.s (flux of 2.0 -- 10.0 keV, absorption
corrected)\\
\end{table}  

\begin{table}
\caption[]{Two Gaussians fits for the Fe K$\alpha$ line of the five time
intervals. The parameters are central energy (E$_c$), width ($\sigma$) and
equivalent width (EW) for two Gaussians. The equivalent widths are 
calculated relative to the X-ray continuum at 6.4 keV}
\begin{tabular}{cccccccccc}
\\ \hline \hline
intervals & E$_c$1 (keV) & $\sigma$1 (keV) & EW1 (eV) 
& E$_c$2 (keV)
 & $\sigma$2 (keV) & EW2 (eV) & $\chi^2$/dof \\\hline
SIS & 6.38$_{-0.01}^{+0.01}$ & 0.08$_{-0.03}^{+0.02}$ & 140$_{-17}^{+16}$ 
& 6.03$_{-0.09}^{+0.10}$ & 0.87$_{-0.08}^{+0.09}$ & 374$_{-34}^{+36}$ 
& 1198/1118 \\
I-1 & 6.39$_{-0.05}^{+0.05}$ & 0.15$_{-0.06}^{+0.06}$ & 164$_{-52}^{+69}$ 
& 5.89$_{-0.31}^{+0.22}$ & 0.95$_{-0.20}^{+0.21}$ & 484$_{-91}^{+100}$ 
& 965/889 \\
I-2 & 6.39$_{-0.03}^{+0.03}$ & 0.00$_{-0.00}^{+0.09}$ & 117$_{-51}^{+25}$ 
& 6.36$_{-0.23}^{+0.26}$ & 0.36$_{-0.15}^{+0.39}$ & 149$_{-69}^{+60}$ 
& 914/858 \\
I-3 & 6.40$_{-0.02}^{+0.02}$ & 0.10$_{-0.04}^{+0.03}$ & 215$_{-56}^{+37}$ 
& 5.51$_{-0.10}^{+0.10}$ & 0.44$_{-0.09}^{+0.11}$ & 220$_{-41}^{+40}$ 
& 927/917 \\
I-4 & 6.37$_{-0.03}^{+0.03}$ & 0.12$_{-0.12}^{+0.04}$ & 166$_{-27}^{+27}$ 
& 5.40$_{-0.17}^{+0.13}$ & 0.33$_{-0.15}^{+0.23}$ & 83$_{-30}^{+53}$ 
& 840/925 \\
I-5 & 6.35$_{-0.03}^{+0.02}$ & 0.00$_{-0.00}^{+0.07}$ & 121$_{-25}^{+32}$ 
& 6.23$_{-0.18}^{+0.18}$ & 0.87$_{-0.16}^{+0.20}$ & 401$_{-71}^{+72}$ 
& 856/910 \\

\hline
\\
\end{tabular}
\end{table}

\begin{table}
\caption[]{Disk line fits for the Fe K$\alpha$ line of the five time
intervals}
\begin{tabular}{ccccccccccc}
\\ \hline \hline
intervals & R$_i$ (R$_g$) & R$_o$ (R$_g$) & EW(eV) & q & $\chi^2$/dof \\\hline

SIS & 8.8$_{-0.9}^{+0.8}$ & 1000$^f$ & 304$_{-17}^{+17}$ 
  & -4.4$_{-0.7}^{+0.9}$ & 1201/1119 \\
I-1 & 7.4$_{-0.6}^{+0.8}$ & 1000$^f$ & 375$_{-41}^{+42}$ 
  & -4.2$_{-0.8}^{+0.6}$ & 975/892 \\
I-2 & 16.8$_{-10.7}^{+45.4}$ & 1000$^f$ & 148$_{-40}^{+44}$ 
  & -2.3$_{-0.9}^{+0.6}$ & 918/861 \\
I-3 & 10.8$_{-4.8}^{+1.7}$ & 18.6$_{-2.9}^{+3.8}$ & 330$_{-37}^{+37}$ 
  & -2$^f$ & 917/920 \\
I-4 & 9.1$_{-3.1}^{+3.4}$ &18.5$_{-5.2}^{+9.2}$ & 171$_{-31}^{+31}$ 
  & -2$^f$ & 839/928 \\
I-5 & 8.3$_{-1.1}^{+1.1}$ & 1000$^f$ & 300$_{-33}^{+33}$ 
  & -4.3$_{-2.6}^{+1.0}$ & 871/913 \\
\hline
\\
\end{tabular}\\
$^f$ frozen because the fit can not give sufficient constraints 
on the parameters.
\end{table}

\end{document}